\documentclass{article}
\usepackage{spconf,amsmath,graphicx}
\usepackage{amssymb}
\usepackage{booktabs}
\usepackage[T1]{fontenc}
\usepackage[utf8]{inputenc}
\usepackage[misc]{ifsym}
\usepackage[colorlinks,linkcolor=red,anchorcolor=blue,citecolor=green]{hyperref}

\name{Xuhua Ren$^\#$, Jiayu Huo$^\#$, Kai Xuan, Dongming Wei, Lichi Zhang$^\ast$, Qian Wang\thanks{Equal Contribution: X.Ren and J.Huo. Corresponding Author: L.Zhang (lichizhang@sjtu.edu.cn). Research was supported by STCSM (17411953300, 19PJ1406800) and Medical-Engineering Cross Research Foundation of Shanghai Jiao Tong University.}}
\address{Institute for Medical Imaging Technology, School of Biomedical Engineering\\
  Shanghai Jiao Tong University\\
  Shanghai, China 200030}

\title{ROBUST BRAIN MAGNETIC RESONANCE IMAGE SEGMENTATION FOR HYDROCEPHALUS PATIENTS: HARD AND SOFT ATTENTION}

\begin{document}
%
\maketitle
\begin{abstract}
Brain magnetic resonance (MR) segmentation for hydrocephalus patients is considered as a challenging work. Encoding the variation of the brain anatomical structures from different individuals cannot be easily achieved. The task becomes even more difficult especially when the image data from hydrocephalus patients are considered, which often have large deformations and differ significantly from the normal subjects. Here, we propose a novel strategy with hard and soft attention modules to solve the segmentation problems for hydrocephalus MR images. Our main contributions are three-fold: 1) the hard-attention module generates coarse segmentation map using multi-atlas-based method and the VoxelMorph tool, which guides subsequent segmentation process and improves its robustness; 2) the soft-attention module incorporates position attention to capture precise context information, which further improves the segmentation accuracy; 3) we validate our method by segmenting insula, thalamus and many other regions-of-interests (ROIs) that are critical to quantify brain MR images of hydrocephalus patients in real clinical scenario. The proposed method achieves much improved robustness and accuracy when segmenting all 17 consciousness-related ROIs with high variations for different subjects. To the best of our knowledge, this is the first work to employ deep learning for solving the brain segmentation problems of hydrocephalus patients.
\end{abstract}
\section{Introduction}
\label{sec:intro}

Hydrocephalus is a condition that there is an abnormal accumulation of cerebrospinal fluid (CSF) in cavities within the patient’s brain. It is related to the introduction of blood and proteins into the CSF during surgery, accident, etc., the change in ventricular size is quite large by comparison of coregistered images. Thus, it is desirable to identify the incidence of the phenomenon, and to correlate the presence or absence of such a phenomenon with clinical symptoms. In order to precisely evaluate the damage of the hydrocephalus, it is important to identify and parcellate the consciousness-related regions-of-interest (ROIs) in brain MR images. However, manual segmentation on these ROIs is generally time-consuming and error-prone due to high inter- or intra-operator variability especially in hydrocephalus patient data. Therefore, fully automated method is essential for hydrocephalus patient brain segmentation.

\begin{figure*}[htb]
\centering
\includegraphics[height=9.5cm,width=0.87\textwidth]{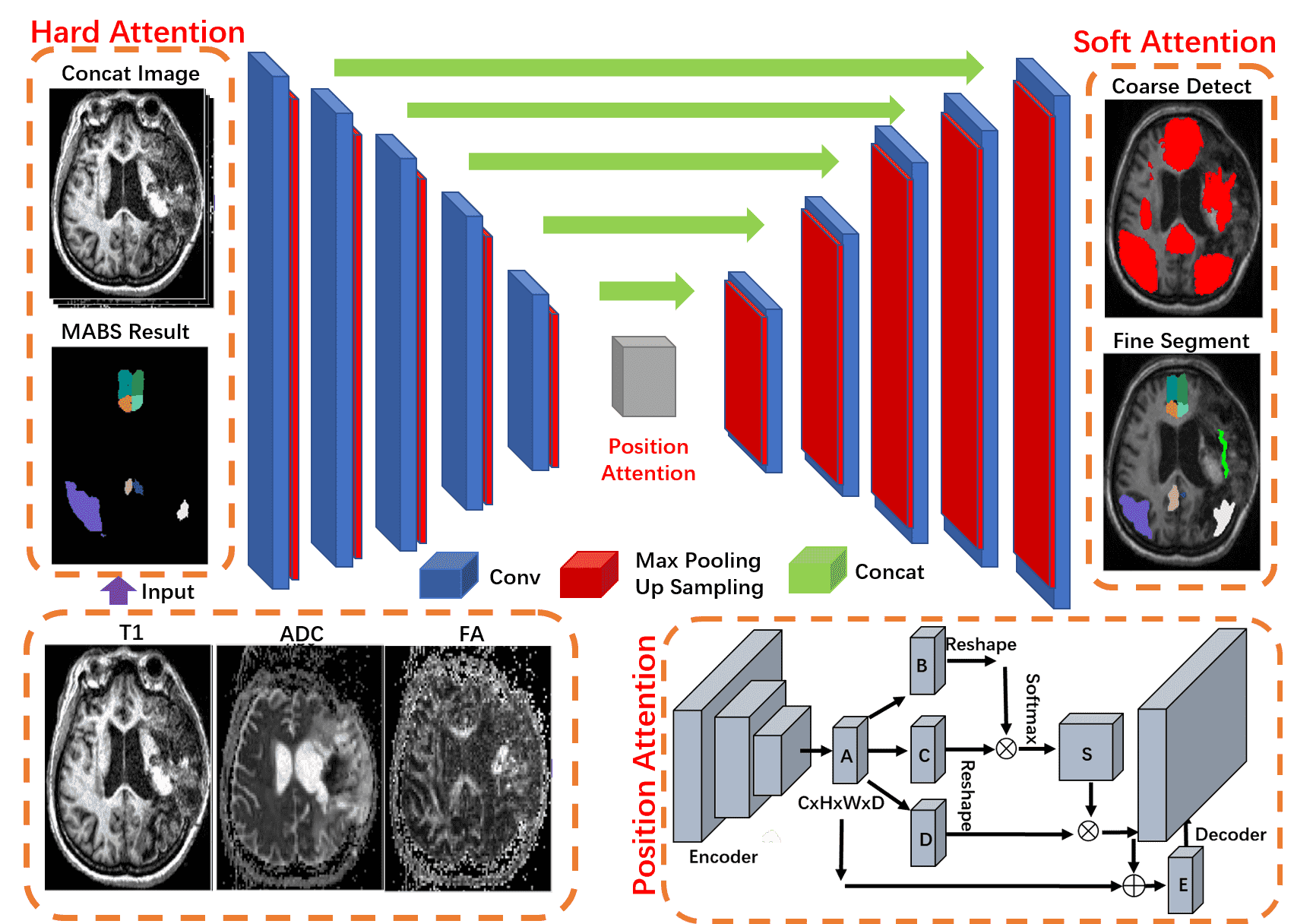}
\caption{Illustration of our proposed method: 1) hard-attention module which combines atlas map from VoxelMorph-based MABS to increase the robustness of model; 2) soft-attention module which decomposes the single segmentation task into several sub-tasks including coarse detection and fine segmentation. Moreover, we address a position-attention module into encoder-decoder architecture for generating spatially long-range contextual information.}
\label{fig1}
\end{figure*}

With the success of deep learning in medical imaging, supervised segmentation approaches built on 3D convolution neural networks (CNNs) have produced accurate segmentation results at high speed. For example, Snehashis \textit{et al.} \cite{roy} segmented white matter lesions from multi-contrast MR images by CNN. Pim \textit{et al.} \cite{moesk} used an adversarial training approach to improve CNN-based brain MR image segmentation. Moreover, Mohsen \textit{et al.} \cite{gha} trained a CNN on legacy brain MR images, and evaluated the performance of the domain-adapted network on the same task but with images from different domains.

On the other hand, the multi-atlas-based segmentation (MABS) is also considered as an important and effective method, especially before the era of deep learning. The process of deformable image registration plays pivotal important role in MABS, which aims at obtaining a deformation field to align the moving image with the fixed image in a topology-preserving way. Conventionally many registration approaches \cite{baj,rue,thi} intend to measure the deformation fields for the to-be-registered image pair in iterative optimization, which results in very slow computation speed. Recent deep-learning-based algorithms, such as VoxelMorph \cite{voxelmorph}, turn to deep neural networks to estimate the deformation fields, and greatly reduces the algorithm runtime with GPU support. 

Although the above-mentioned segmentation methods are effective for normal brain images, segmenting the hydrocephalus patients' images remains a challenging task, due to the high variations of anatomical structures. Specifically, the pathology-induced deformations vary a lot between individuals. Traditional CNN architectures generalize poorly on distorted samples such as hydrocephalus data. Therefore, our goal is to resolve the segmentation problem when using hydrocephalus patients training data by CNN architecture model, which is further described as follows. 

In this paper, we propose a novel CNN-based method for automatic brain MR image segmentation, the main contributions are three-fold: \textbf{$1)$} We deploy the hard-attention module for generating atlas map, which includes some crucial ROIs that can resist the influences from large variations of anatomical structures as well as limited training samples. We also utilize VoxelMorph to significantly speed up MABS algorithm and generate segmentation results as our attention maps. \textbf{$2)$} We refer from DANet \cite{danet} and develop a multi-level soft attention module for better generalization on distorted samples, which splits the brain into two tasks, i.e. coarse detection and fine segmentation, according to anatomical knowledge. We also address the position self-attention mechanism to capture the shape variations between any two regions of the feature maps, to ease the poor generalization abilities on distortion samples. \textbf{$3)$} The proposed method achieves much improvement in robustness and accuracy dealing with all 17 ROIs of large variations. To the best of our knowledge, this is the first work that employ deep learning for hydrocephalus patient brain segmentation.

\section{material and method}
\label{sec:typestyle}

\subsection{Material}
\label{submat}
In this work, we collected 21 clinical hydrocephalus patients, each of which obtained the MR images in T1, FA and ADC modalities. Note that all these subjects have hematoma volume and hydrocephalus disease, causing large deformation in brain anatomical structures. Specifically, our aim is to segment the consciousness-related brain regions on 3T MRI scans following \cite{wu18}, which are Insula-R (IR), Insula-L (IL), Thalamus-R (TR), Thalamus-L (TL), internal Capsule-R-Ant (ICRA), internal Capsule-R-Post (ICRP), internal Capsule-L-Ant (ICLA), internal Capsule-L-Post (ICLP), Cingulate-R-Ant (CRA), Cingulate-R-Post (CRP), Cingulate-L-Ant (CLA), Cingulate-L-Post (CLP), Medial prefrontal cortex-R (MCR), Medial prefrontal cortex-L (MCL), Inferior parietal lobule-L (IPL), Inferior parietal lobule-R (IPR) and Brainstem (B). 

Note that we grouped the annotated images into five-fold cross-validation sets. Some preprocessing works have also been made in order to feed the neural network with this data:  First, the images were resized to make them isotropic, with a voxel size of $1 \text{mm} \times 1 \text{mm} \times 1 \text{mm}$ with trilinear interpolation. Then, the images were normalized (dividing by the maximum intensity value in foreground region) in order to improve convergence. We only utilized random flip in three directions during training and we did't use any augmentation during testing period.

There are two modules designed for our hydrocephalus brain segmentation method: the hard and soft attention modules. The whole pipeline of our method and proposed hard and soft attention fully convolutional network (FCN) is shown in Fig.~\ref{fig1}. The details about the two modules are further illustrated in Section 2.1 and 2.2, respectively.

\subsection{Hard attention module}
\label{subhard}

The hard attention module is designed for generating attention maps as prior knowledge.
In our implementation, the coarse segmentation results obtained with MABS
are taken as hard attention maps,
and we use VoxelMorph for more efficient and robust deformable registration.
More specifically, we learn parameters $\theta$ of a function $g_\theta$ which will generate deformation field $\phi$ to warp a 3D volume $m$ to a fixed volume $f$. The loss function is calculated by measuring negative similarity between the warped image $\phi \cdot m$ and the fixed image $f$.
During test stage, given the new images $f$ and $m$, the deformation field is obtained by evaluating $\phi = g_\theta(f, m)$ and the segmentation for $m$ is transferred through warping the label of $f$ with $\phi$.
We have multiple segmentation results for $m$ with different fixed images $f$, and these transferred labels will be further fused to a single consensus segmentation, which are taken as our hard attention map. Note that there have no data leakage in this module.

\begin{figure*}[htb]
\centering
\includegraphics[height=7cm,width=\textwidth]{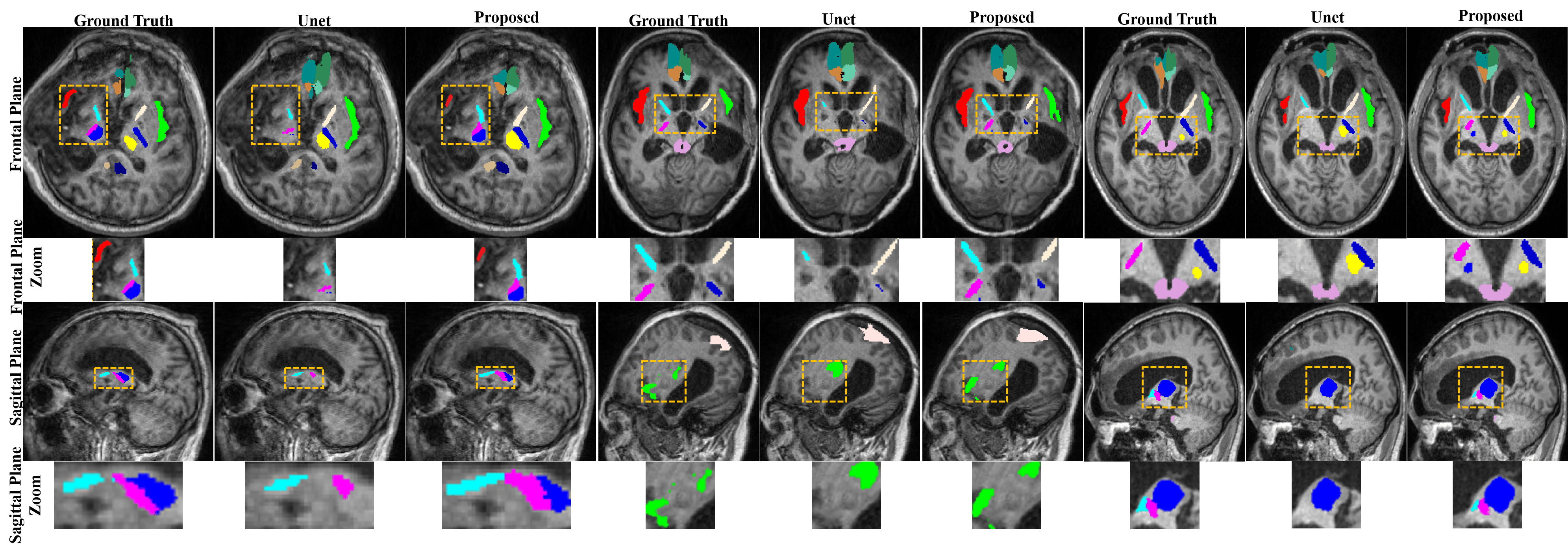}
\caption{Visualization of segmentation results for hydrocephalus patients.}
\label{fig2}
\end{figure*}

\subsection{Soft attention module}
\label{subsoft}

The multi-level soft-attention module consists of the following major processes: First, we split all of the ROIs shown in Fig.~\ref{fig1} into two groups based on the anatomical knowledge. We first merge the whole ROIs as foreground, and utilize the first classifier to local the foreground to ease the task difficulty. The cropped region of the input images based on the first classifier result is used as the input of the second classifier for 17 ROIs fine-grained segmentation.

Second, we design a position attention module in encoder-decoder FCN. We feed the encoder features into the position attention module, and generate new features of spatial long-range contextual information through the following three steps: 

\begin{enumerate}
\item Generate a spatial attention matrix which models the spatial relationship between any two pixels of the features;
\item Perform a matrix multiplication between the attention matrix and the original features;
\item Perform an element-wise sum operation on the above multiplied resulting matrix and original features to obtain the final representations reflecting long range contexts.
\end{enumerate}

\begin{table}
\centering
\caption{Comparisons with state-of-the-arts and ablation studies}
\begin{tabular}{cc}
\hline
Model& Dice(\%)\\
\hline
\multicolumn{2}{l}{Ablation studies}\\
\hline
Base& $61.34\pm 20.05$\\
Base + Hard& $65.64\pm 18.35$\\
Base + Hard + Soft& \boldmath{$67.19\pm 17.18$}\\
\hline
\multicolumn{2}{l}{Comparison with the state-of-the-art methods}\\
\hline
Unet \cite{3dunet}& $61.34\pm 20.05$\\
Vnet \cite{vnet}& $61.55\pm 20.45$\\
PSP \cite{psp}& $61.81\pm 20.01$\\
ENC \cite{enc}& $61.90\pm 19.93$\\
DEEP \cite{deeplab}& $62.05\pm 19.63$\\
Base + Hard + Soft& \boldmath{$67.19\pm 17.18$}\\
\hline
\end{tabular}
\label{table1}
\end{table}

\begin{table*}
\centering
\tiny
\caption{Comparison of the proposed method with state-of-arts}
\begin{tabular}{ccccccccc}
\hline
& IR& IL& TR& TL& ICRA& IPR& ICRP& ICLA\\
\hline
Unet& $60.32\pm 28.14$& $62.16\pm 24.08$& $65.58\pm 25.18$& $71.69\pm 18.21$& $65.87\pm 19.14$& $50.80\pm 20.01$& $64.19\pm 17.23$& $56.46\pm 18.83$\\
Vnet& $61.25\pm 30.31$& $61.02\pm 26.69$& $66.12\pm 28.96$& $71.43\pm 14.38$& $67.58\pm 17.46$& $52.49\pm 18.73$& $65.30\pm 19.49$& $56.19\pm 19.28$\\
PSP& $61.45\pm 27.37$& $62.74\pm 23.33$& $66.56\pm 23.99$& $72.87\pm 18.17$& $65.48\pm 18.75$& $51.71\pm 17.43$& $64.26\pm 16.99$& $57.15\pm 20.43$\\
ENC& $60.02\pm 28.11$& $63.08\pm 22.57$& $66.73\pm 25.04$& $71.62\pm 17.54$& $65.72\pm 17.42$& $50.87\pm 21.49$& $65.04\pm 16.52$& $57.16\pm 17.03$\\
DEEP& $60.98\pm 28.08$& $63.57\pm 23.24$& $66.65\pm 26.15$& $71.43\pm 18.76$& $67.54\pm 19.49$& $50.75\pm 20.21$& $65.01\pm 16.32$& $56.61\pm 20.43$\\
Prop& \boldmath{$66.43\pm 25.67$}& \boldmath{$69.73\pm 23.29$}& \boldmath{$71.41\pm 21.54$}& \boldmath{$75.77\pm 11.41$}& \boldmath{$68.51\pm 20.19$}& \boldmath{$55.69\pm 19.72$}& \boldmath{$69.32\pm 16.81$}& \boldmath{$61.89\pm 17.16$}\\
\hline
& CRA& CRP& IPL& CLA& CLP& MCR& MCL& B\\
\hline
Unet& $58.52\pm 13.19$& $48.09\pm 21.33$& $60.98\pm 11.48$& $63.08\pm 19.91$& $62.65\pm 16.86$& $56.70\pm 24.97$& $54.58\pm 24.93$& $88.40\pm 13.89$\\
Vnet& $59.14\pm 13.78$& $48.90\pm 20.77$& $59.57\pm 9.08$& $63.12\pm 22.55$& $65.03\pm 12.42$& $56.62\pm 16.71$& $56.31\pm 21.49$& $90.23\pm 8.41$\\
PSP& $58.79\pm 10.30$& $48.84\pm 23.26$& $61.23\pm 9.30$& $63.75\pm 21.86$& $62.10\pm 16.69$& $56.77\pm 25.43$& $55.44\pm 26.81$& $88.10\pm 15.18$\\
ENC& $59.26\pm 14.15$& $48.95\pm 22.18$& $61.45\pm 11.98$& $62.85\pm 20.78$& $63.63\pm 16.20$& $58.35\pm 24.31$& $55.44\pm 24.91$& $90.11\pm 13.77$\\
DEEP& $58.63\pm 13.96$& $49.10\pm 20.43$& $60.57\pm 12.35$& $63.81\pm 17.69$& $63.78\pm 14.87$& $58.18\pm 22.65$& $55.66\pm 22.62$& $89.98\pm 12.38$\\
Prop& \boldmath{$61.86\pm 15.98$}& \boldmath{$58.12\pm 13.34$}& \boldmath{$68.39\pm 10.58$}& \boldmath{$68.91\pm 18.06$}& \boldmath{$69.29\pm 10.40$}& \boldmath{$64.97\pm 22.59$}& \boldmath{$58.62\pm 21.81$}& \boldmath{$92.38\pm 3.37$}\\
\hline
\end{tabular}
\label{table2}
\end{table*}

As in Fig.~\ref{fig1}, given a local feature $A$, we first feed it into a convolution layers to generate two new feature maps $B$ and $C$, respectively, where $\{B, C\} \in \mathbb{R}^{H \times W} $. Then, we reshape them to $\mathbb{R}^{N}$, where $N = H \times W$ is the number of pixels. After that, we perform a matrix multiplication between the transpose of $C$ and $B$, and apply a softmax layer to calculate the spatial attention map $S$. Meanwhile, we feed feature $A$ into a convolution layer to generate a new feature map $D \in \mathbb{R}^{H \times W} $, and reshape it to $\mathbb{R}^{N}$. We also perform a matrix multiplication between $D$ and the transpose of $S$, and reshape the result to $\mathbb{R}^{N}$. Finally, we conduct an element-wise sum operation with the features $A$ to obtain the final output $E$.

\section{experiments}
\label{sec:experment}

In this section, we detail the results of our proposed automatic brain segmentation method in hydrocephalus dataset including hard- and soft-attention, which is also compared with the alternatives. The network is trained and applied with a Titan X GPU on Tensorflow and NiftyNet platform. Regarding the hyper-parameters, the basic learning rate is set to 0.0001. For multi-task learning, the learning rate decreases gradually. The momentum and weight decay are set to 0.9 and 0.0001, respectively. Note that we only utilize random flipping in three directions during training.

Table \ref{table1} shows the robustness of hard attention module in experiments. Generally, when concatenating atlas map, the MABS segmentation result with the original volumes feeding to the FCN model ("Base") achieves higher accuracy on the hydrocephalus patient dataset ("Base" vs "Base + Hard"). According to the result, this atlas map can give prior knowledge for the current segmentation task, which improve the segmentation accuracy and robustness. 

To validate our framework with multi-level soft-attention module, we also compare several different settings and report the results in Table \ref{table1}. First, we consider the single task of segmentation only. The Dice score of our implementation is $61.34\pm 20.05$. Second, with the network architecture ("Base + Hard" vs "Base + Hard + Soft") validated in multi-task learning, we further verify the contribution of the proposed soft-attention module. The experimental results in the middle of Table \ref{table1} shows it could outperform other solutions. Therefore, we conclude that the proposed soft-attention is beneficial to the segmentation task.

Finally, we compare our proposed method with other state-of-the-art algorithms including Unet \cite{3dunet}, Vnet \cite{vnet}, PSPnet \cite{psp}, Encnet \cite{enc} and DeeplabV3 \cite{deeplab} in Table \ref{table1}. The results show that the proposed method outperforms all the methods under comparison in the five-fold validation set. We have also provided visual inspection of the typical segmentation results (Unet vs. proposed method) with the ground truth in Fig.~\ref{fig2}. The labeling result of the region inside the yellow box shows that, with the integration of our proposed module, the labeling accuracy and robustness is improved. Four patient cases which are failed to be located by Unet are successfully captured by our framework. 

Furthermore, we compare our method with the state-of-arts method for each ROI in Table \ref{table2}, it is shown that our method has demonstrated top-tier performance in each of ROIs compared with Unet solution. Our method can obtain more accurate result for hydrocephalus patient dataset. 

\section{Conclusion}
\label{sec:illust}

We have proposed a novel and effective atlas-guided, multi-level soft-attention framework for semantic segmentation of brain MR images in hydrocephalus patient dataset. Specifically, we proposed a hard-attention module to give the prior knowledge from VoxelMorph based MABS method to improve the robustness of network. Moreover, we decomposed the very challenging semantic segmentation task to several sub-tasks, which are associated with coarse-to-fine segmentation mechanism. Finally, we addressed the position attention module to FCN model which could capture long-range contextual information in the model. We have conducted comprehensive experiments on a large, custom medical image dataset which is the first work for hydrocephalus patient brain segmentation task according to our knowledge.



\bibliographystyle{IEEEbib}
\bibliography{refs}
\end{document}